\documentclass[twoside,a4paper,11pt]{sea22}
\usepackage{graphicx}
\usepackage{hyperref}
\usepackage{movie15}
\usepackage{color}
\begin{document}
\pagenumbering{arabic}
\pagestyle{myheadings}
\thispagestyle{empty}
{\flushleft\includegraphics[width=\textwidth,bb=58 650 590 680]{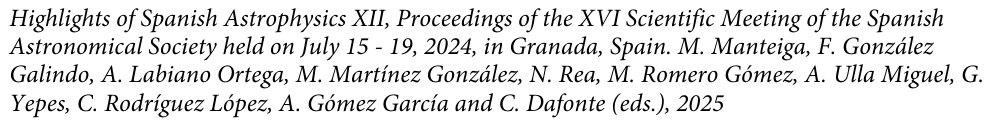}}
\vspace*{0.2cm}
\begin{flushleft}
{\bf {\LARGE
%
%%% TITLE of the paper. 
%%% TITLE of the paper. 
\textcolor[rgb]{0,0,1.0}{A}strophysical \textcolor[rgb]{0,0,1.0}{T}ransients in \textcolor[rgb]{0,0,1.0}{T}ime \textcolor[rgb]{0,0,1.0}{D}omain and \textcolor[rgb]{0,0,1.0}{M}ulti-\textcolor[rgb]{0,0,1.0}{M}essenger \textcolor[rgb]{0,0,1.0}{A}stronomy: a review
%
% Do not delete next few lines
}\\
\vspace*{1cm}
%
%%% Include here the LIST OF AUTHORS.
%%% Include here the LIST OF AUTHORS.
%%% Note that the last author has to be preceeded by an AND.
Mar\'{\i}a D. Caballero-Garc\'{\i}a$^1$ 
%
% Do not delete next few lines
}\\
\vspace*{0.5cm}
%
%%% AFFILIATIONS LIST.
%%% and the AFFILIATIONS LIST. Note that one affiliation per line.
%%% Add as many affiliations as necessary. 
$^1$ 
Instituto de Astrof\'isica de Andaluc\'ia (IAA-CSIC), Glorieta de la Astronom\'ia s/n, E-18008, Granada (Spain)\\
%
% Do not delete next few lines
\end{flushleft}
%
% Headings
\markboth{
%%% Type the SHORT version of the paper title.
%%% Type the SHORT version of the paper title.
Astrophysical transients in time domain and multi-messenger astronomy
}{ % Do not delete
%
%%%  First Author \& Second Author   OR   First-author et al. 
%%%  First Author \& Second Author   OR   First-author et al. if the author list 
%%% contains three or more authors.
Caballero-Garc\'{\i}a, M.
% 
% Do not delete next few lines
}
\thispagestyle{empty}
\vspace*{0.4cm}
\begin{minipage}[l]{0.09\textwidth}
\ 
\end{minipage}
\begin{minipage}[r]{0.9\textwidth}
\vspace{1cm}
\section*{Abstract}{\small
%
% ABSTRACT ABSTRACT ABSTRACT
% ABSTRACT ABSTRACT ABSTRACT
%%% Type the ABSTRACT of your paper
The wide fields of view, high sensitivity, and broad energy coverage of current X-ray and gamma-ray satellites, coupled with the high cadence observational strategy of some of them (recently {\it Swift} and {\it Fermi}) have been ideal for carrying out 
unprecedented studies of the variability properties of different classes of Galactic and extra-Galactic high-energy sources.

These classes of objects range from nearby flaring stars to the most distant Active Galactic Nuclei (AGNs). In this paper we focus on some of the most energetic events, i.e. those powered by accretion (black hole binaries, ultra-luminous X-rays and 
Active Galactic Nuclei) until those leading to the first detections of Gravitational Waves (GWs), i.e. the Gamma-ray Bursts (GRBs), passing through the controversial Inter-Mediate Mass Black-Holes (IMBHs). 

We show the importance of X-ray and gamma-ray emission for the determination of the properties (mass, spin, inclination, height of the corona) of the compact objects residing in systems powered by accretion and the role of the LIGO and VIRGO 
interferometers in the case of the (compact objects) binary mergers. Gravitational waves allow the determination of the properties of the non-light-emitting compact objects (black-holes; BHs) in binary mergers (BH-BH) that otherwise would be nearly impossible.

We show that for the case of neutron star (NS) mergers (NS-BH or NS-NS) electro-magnetic (EM) emission is still possible, and very powerful, being the responsible for the X-ray and gamma-ray emission of many GRBs (kilonovae). Very recently these transients have 
been discovered to show X-ray and gamma-ray variability patterns that could lead to very important insights into the properties of the binary mergers that originated them, opening a new view for their study.

%
% Do not delete next few lines
\normalsize}
\end{minipage}
%
%
%%% BODY of the paper
%%% BODY of the paper
%
\section{Introduction}

$\,\!$\noindent Transient sources have extensively been discovered through observations in the whole electro-magnetic (EM) domain. Depending on the nature of the physical process involved the emission occurs in a particular energy range. Usually, for the transients
associated to the most powerful phenomena the energy range involved increases, with the X-ray and gamma-ray emission reserved to the most powerful phenomena. This is what happens in the death of the most massive stars, leading to the formation of
a compact object, either a Neutron Star (NS) or a Black-Hole (BH). In the most powerful events, a.k.a. Gamma-Ray Bursts, the compact objects remain ``isolated", but leaving behind a very complex phenomenology sometimes
accompanied with emission in the form of Gravitational Waves (GWs). In the case of less powerful events, alike the death of less-massive stars via Super-Novae explosions, i.e. SNe) in binary systems, the two components of the binary system persist in orbit with each other. Usually an accretion
disc is formed (most of the times around the compact object, with the companion star being the donor) constituting an X-ray BH or a NS-binary system. Although the
(relative) simplicity of the former (BHBs), the understanding of their emission phenomenology serves as the prototype for the emission of the rest of compact objects, with little variations of it (\cite{homan05,caballero07,caballero09,caballero13b} and references therein).

The mass of the BH and/or NSs in binary systems can be derived through the properties of their X-ray emission. X-ray timing and spectra are usually studied to understand the close vicinity of them, thus providing the best signals
of the compact object (under the application of some hypothesis/models to the data). The most robust way of deciphering the nature of the components of a binary system is through their radial velocities but they are limited to 
the most nearby cases and with the biggest optical telescopes available. Therefore, X-rays and (soft) gamma-rays remain many times as the only way to unveil the nature of the compact objects of the (most) powerful transients populating the Universe (see e.g. \cite{caballero18,caballero20,alston20} and references therein).

In this paper we present an overview in the context of the findings from the new multi-messenger era so far (Sec. \ref{multim}). We end the review including observations of the compact objects residing the gamma-ray 
bursts (GRBs; Sec. \ref{grbs}) thus including the case of very magnetized NSs (Sec. \ref{magnetars}) and future perspectives (Sec. \ref{future}).

\section{The multi-messenger astronomy}   \label{multim}

$\,\!$\noindent Before 2016 it was thought that there were two classes of black holes: stellar-mass black holes, with masses no more than ${\approx}10\,{\rm M}_{\odot}$                      
and super-massive black holes at the centre of galaxies with masses in the range of $10^{4}-10^{9}\,{\rm M}_{\odot}$. It was the recent birth of Gravitational Wave Astronomy what has allowed the discovery of the first sample of the Intermediate-Mass Black Holes (IMBHs,${\rm M}=20-100\,{\rm M}_{\odot}$) by the Advanced Laser Interferometer Gravitational-wave Observatory (LIGO) and VIRGO interferometer observations. 

These IMBHs are detected through the GWs emitted by binary BH mergers, so there is little expectation for the detection of an EM counterpart (even for the most sensitive current instrumentation). Because BHs do not emit light their presence
can only be inferred through the gravitational effect on their neighbours (e.g. Tidal Disruption Events; TDEs) or due to the light (often in X-rays) emitted by the hot gas falling onto them (i.e. the so-called BHBs, mentioned above). Therefore, BH mergers can
only be detected through their GWs \footnote{BHs can be detected through GWs just in the case they are accelerated.}.

\begin{figure}
\center
\includegraphics[scale=0.5,angle=270]{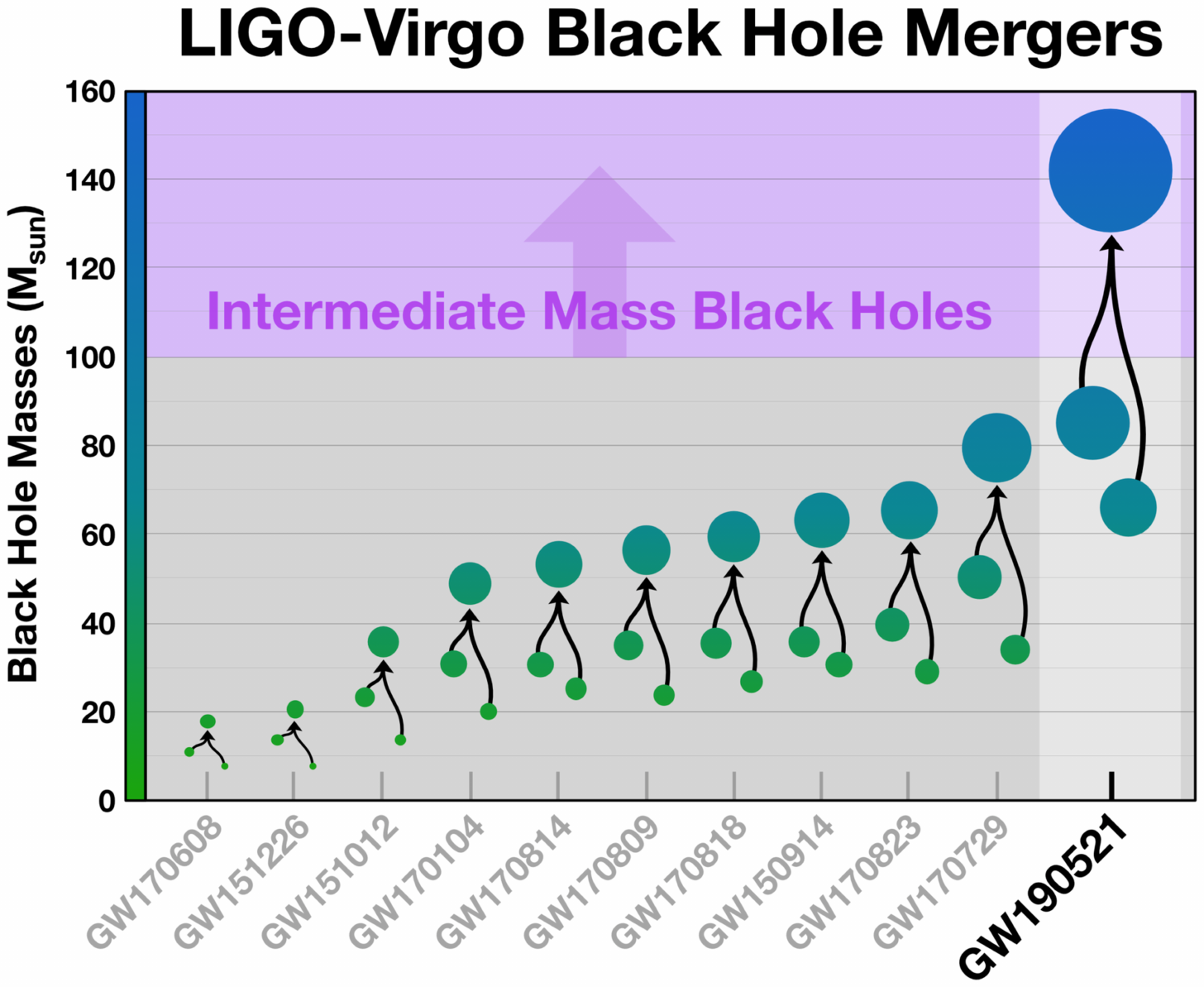}
\caption{\label{fig5} LIGO and Virgo have observed their largest black hole merger to date, an event called GW190521, in which a final black hole of 142 solar masses was produced. This chart compares the event to others witnessed by LIGO and Virgo and indicates that the remnant of the GW190521 merger falls into a category known as an intermediate-mass black hole – and is the first clear detection of a black hole of this type. Intermediate-mass black holes, which have previously been predicted theoretically, would have masses between those of stellar-mass black holes and the super-massive ones at the hearts of galaxies. 
[Image credit: : LIGO/Caltech/MIT/R. Hurt (IPAC)].
}
\end{figure}

IMBHs in the mass range of $30-130\,{\rm M}_{\odot}$ (as those found by the interferometers; Fig.~\ref{fig5}) are formed through BH-BH mergers in low-metallicity environments. It has been proposed that they could form from the 
unseen dark matter proposed to exist in the galactic haloes. On the other hand IMBHs may be produced in runaway mergers in the cores of young
clusters. Nevertheless, there is still no firm detection of their EM counterpart (only a handful of detection candidates) that would give more clues about their existence and origin.

\subsection{The Gamma-Ray Bursts}   \label{grbs}

$\,\!$\noindent In Sec. \ref{multim} we have focused in the case of IMBHs but until their presence is fully understood there is still a lot to tell about EM counterparts of GW events already discovered. Indeed Gamma-Ray Bursts (GRBs)
are the only confirmed sources known to be emitting both in the EM spectrum and in GWs. 

GRBs appear as brief flashes of high-energy photons, emitting the bulk of their energy above 0.1 MeV and becoming recognised as the most energetic phenomena in the Universe after the Big Bang. They are
(mainly) classified into two categories depending on their duration: short GRBs (sGRB; ${\le}2$\,s) and long GRBs (lGRB; $>2$\,s). lGRBs are associated with supernovae (SNe) so they are believed to 
originate from massive core-collapse star events. On the other hand sGRBs were thought to be related to compact star mergers and henceforth were expected to be accompanied by kilonovae.

The detection of GWs from a coalescing black hole binary system has been one of the major discoveries in this 21st century so far. Furthermore, the detection of the first GW 
EM counterpart (GW~170817) opened the new era of multi-messenger Astronomy in 2017 (see \cite{hu23} for a review). For the first time, this shed light into three open issues that remained obscure until that time: 1) electromagnetic 
counterparts can be detected for at least a fraction of GW alerts related to neutron stars (NS-NS) mergers; 2) short duration gamma-ray bursts arise in these NS-NS mergers; and 3) heavy elements (heavier 
than Fe) are produced in these sites due to the r-process nucleosynthesis (i.e. kilonovae). Indeed it is believed that sGRBs are powered by the accretion of a massive remnant
disc onto the compact BH or NS formed following the compact binary merger. But this finding should not be limited to sGRBs since recently (\cite{lu22,troja22}) it has been discovered 
that some lGRBs have kilonovae properties and therefore constitute candidates to GW emission as well.

\begin{figure}
\center
\includegraphics[scale=0.5]{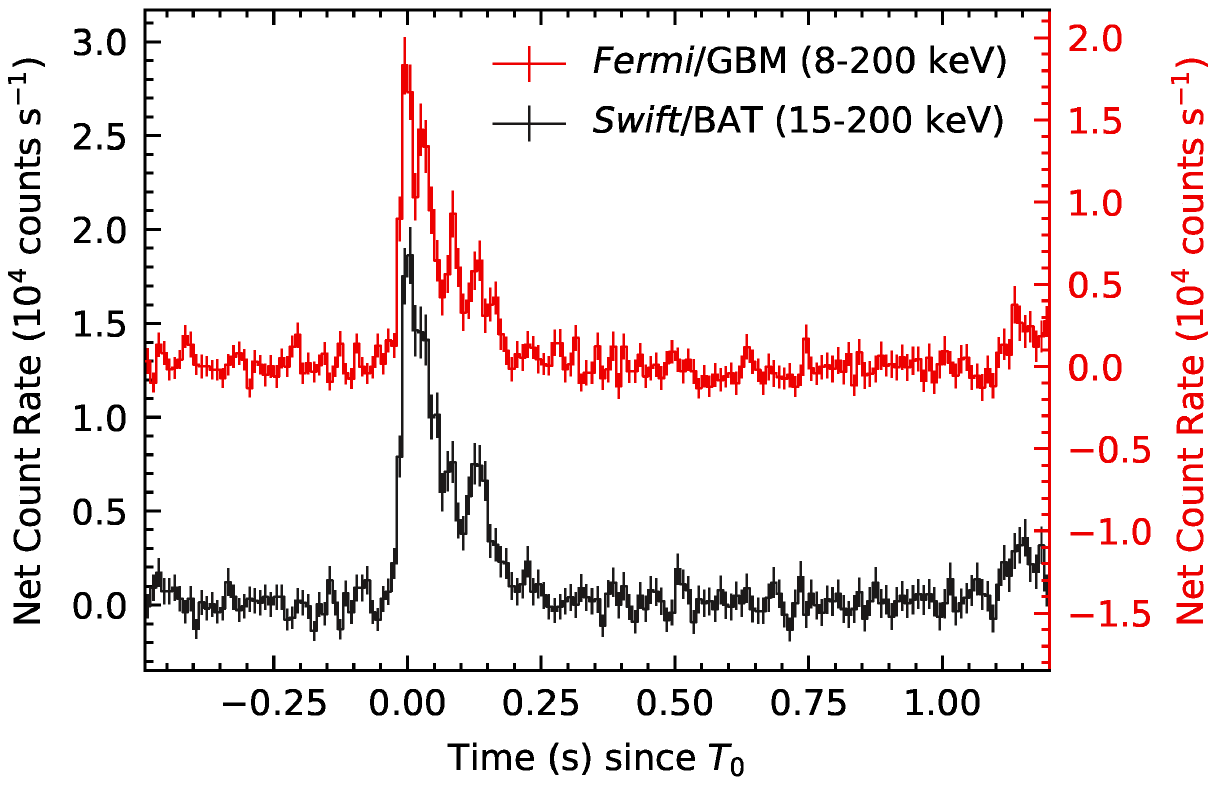}
\caption{\label{fig6} The net lightcurves of GRB 211211A from \textit{Fermi}/GBM and \textit{Swift}/BAT. These two curves are found to consist of some (quasi-)periodic pulses near $T_0$. 
[Image credit: : Xiao, et al. (2022)]. 
}
\end{figure}

\subsection{The compact object residing in the Gamma-Ray Bursts}  \label{magnetars}

$\,\!$\noindent It has been shown in the literature that the X-ray timing and spectroscopy analysis together is (most of the times) the only way to confirm the nature of the compact object residing in the X-ray binary systems. Something similar 
happens with the case of GRBs (in the cases of no GW detection, that constitutes the majority). For BHs residing as the remnants of GRBs they are difficult to be confirmed and only indirect evidences are used. For example, collapsars
(giving rise to the lGRBs) are the best candidates, in particular the most powerful ones. This is due to the fact that they happen due to the collapse of very massive stars (${\sim}10\,{\rm M}_{\odot}$) that give a BH after their death. But
in the case of the vast majority of GRBs the remnant of the explosions is a NS, many times a magnetar (i.e. extremely highly magnetized NS).

It has been shown that isolated magnetars can be identified throughout the presence of characteristic very high frequency and multiple (harmonic) Quasi-Periodic Oscillations (QPOs) in their (X-ray and gamma-ray) light curves (2132\,Hz, 4250\,Hz; \cite{castro21}). Although several 
scenarios have been proposed for the origin of them the most promising is the one saying that they are due to strong perturbations caused by the radial overtones of the fundamental magneto-elastic oscillations in the crust of the neutron star. Thus they likely appear as a consequence of the presence of 
the crust on the surface of the NS.

In the case of the NS-BH and/or NS-NS mergers it has been proposed that in the late in-spiral phase the tidal force on the magnetar due to its (compact) companion would increase dramatically as the components of the binary approach. The tidally-induced 
deformation may surpass the maximum that the crust of the magnetar could sustain just seconds or sub-seconds before the coalescence. A catastrophic global crust destruction could then occur, and the magnetic energy stored in the interior of 
the magnetar would be released thus being observed as a super-flare with an energy release of hundreds of times larger than the giant flares of magnetars, thus leading to a (short)Gamma-ray Burst. Such mechanism could explain the recently observed precursor of GRB 211211A, including 
its claimed (high-frequency) quasi-periodic oscillation at ${\approx}20$\,Hz (\cite{suvorov22,xiao22,chirenti24}; see Fig.~\ref{fig6}). 

\section{The future}  \label{future}

The {\it Swift}\footnote{\url{https://swift.gsfc.nasa.gov}} telescope has three instruments which work together to provide rapid identification and multi-wavelength follow-up of gamma-ray bursts (GRBs) and their afterglows. The Burst Alert Telescope (BAT; 15-150\,keV) has a large 
FOV (2 stero-radians) and high sensitivity allowing the detection of ${\approx}100$ bursts per year with provided on-board coordinates of arc-minute positional accuracy. The X-ray Telescope (XRT; 0.3-10 \,keV) obtains softer light curves and spectra allowing
higher accuracy position burst localizations. The UV/Optical Telescope (UVOT; 170-600\,nm) allows a 0.5 arcsecond position localization of the burst plus the temporal evolution of the UV/optical afterglow. Spectra can also be taken for the brightest 
UV/optical afterglows, which can then be used to determine the redshift via the observed wavelength of the Lyman-alpha cut-off. The results on-board are publicly distributed within seconds for immediate world-wide follow-up with ground facilities
(optical telescopes and GW interferometers).

The {\it Fermi}\footnote{\url{https://fermi.gsfc.nasa.gov}} Gamma-ray Space Telescope is an international and multi-agency space mission launched on 2008 that studies the sky in the energy range 10\,keV-300\,GeV. It has two instruments on-board, i.e. the Large Area Telescope (LAT; 20\,MeV-300\,GeV) 
and the Gamma-ray Burst Monitor (GBM; 8\,keV-40\,MeV).

Both satellites, together with the ESA {\it XMM-Newton} (see elsewhere in this proceedings) and {\it INTEGRAL} \cite{\cite{winkler03}} have enormously contributed in the field of X-ray and gamma-ray astronomy of transients, basically due to their autonomous analysis on-board and rapid distribution (plus later characterization) of the results worldwide. This allowed 
the study of new transients everywhere in the sky since their very early occurrence (mainly GRBs, SNe, Novae,...). This capacity was not shared by any other X-ray and/or gamma-ray previous facility.

In the future we will need
to monitor the transient Universe in the X-ray and gamma-ray (0.3 keV to 20 MeV) energies together with the infrared (in order to see further and deeper away) with an unprecedented time resolution and large detector area. {\it THESEUS} (proposed to ESA for the M7 call) is addressed to detect, observe, and locate transient high-energy events at very high redshift thus making a valuable contribution to Cosmological, transient, and multi-messenger Astrophysics. The discoveries of this (potential) new satellite could change the paradigm of the Universe 
formation, monitoring an unprecedentedly big area of the sky arriving farther (in X-rays and gamma-rays) than any other high-energy instrument has ever achieved in synergy with the observatories (both spatial and ground) available in the 2030s-2040s.

%
%
% Do not delete the next line
\small  % Do not delete
%
%%% Comment the following line if you do not have acknowledgments.
\section*{Acknowledgments}   % Do not delete if you declare acknowledgments
%
%%% ACKNOWLEDGMENTS
%%% ACKNOWLEDGMENTS
MCG acknowledges financial support from the Spanish Ministry project PID2020-118491GB-I00 funded by MCIN/AEI/10.13039/501100011033 and the Severo Ochoa grant CEX2021-001131-S funded by MCIN/AEI/10.13039/501100011033.
%
% Do not delete the next few lines

%
\end{document}